\documentstyle[algo,epsf,11pt]{article}

\title{{\small\hfill Technical Note HLRZ-95-20\\
              \hfill WUB-95-12}\\[1cm]
Many masses on one Stroke: Economic Computation of Quark Propagators}

\author{Andreas Frommer, Bertold N\"ockel\\ {\small Mathematics
    Department, University of Wuppertal, Wuppertal, Germany}\\[0.5cm]
  Stephan G\"usken, Thomas Lippert, Klaus Schilling\\ {\small Physics
    Department, University of Wuppertal, Wuppertal, Germany}}

\begin{document}

\maketitle
\newcommand{\eqnref}[1]{\mbox{Eq.~\ref{#1}}}
\newcommand{\one}{\mbox{{\bf 1}}}
\newcommand{\C}{\makebox{{\bf C}}}

\begin{abstract}
  The computational effort in the calculation of Wilson fermion quark
  propagators in Lattice Quantum Chromodynamics can be considerably
  reduced by exploiting the Wilson fermion matrix structure in inversion
  algorithms based on the non-symmetric Lanczos process.  We consider
  two such methods: QMR (quasi minimal residual) and BCG (biconjugate
  gradients).

  Based on the decomposition $M/\kappa={\bf 1}/\kappa-D$ of the Wilson
  mass matrix, using QMR,
  one can carry out inversions on a {\em whole} trajectory
  of masses simultaneously, merely at the computational expense of a single
  propagator computation. In other words, one has to compute the
  propagator corresponding to the lightest mass only, while all the
  heavier masses are given for free, at the price of extra storage.

  Moreover, the symmetry $\gamma_5\, M= M^{\dagger}\,\gamma_5$ can be
  used to cut the computational effort in QMR and BCG by a factor of two.
  We show that both methods then become---in the critical regime of small
  quark masses---competitive to BiCGStab and significantly better than
  the standard MR method, with optimal relaxation factor, and CG as
  applied to the normal equations.
\end{abstract}

\section{Introduction} 

Lattice QCD allows to compute physical quantities like the hadronic
spectrum, weak decay constants and other weak matrix elements without
recourse to perturbation theory \cite{MONTVAY}.  Basic building blocks
for the construction of such observables are the quark propagators,
i.e. the Green's functions of the fermionic operator. In practice, the
latter are determined via an iterative inversion method on an ensemble
of gauge background fields, generated in a Monte Carlo process.  Thus,
the evaluation of quark propagators represents a major task within
this branch of elementary particle theory.

One can work in two directions in order to achieve good efficiency
computing the propagators:
\begin{itemize}
\item acceleration of the convergence using improved iterative
  procedures \cite{BF94,FHNLS94},
\item exploitation of the structure of the matrix $M$ in the
  implementation of the inverter.
\end{itemize}
Of course, these two directions are not mutually exclusive.

In this note we go one step into the second direction and point out
how the structure of $M$ can be exploited in iterative methods based
on the non-symmetric Lanczos process \cite{LANCZOS50}. Specific
methods in this class comprise the BCG (biconjugate gradient) method
of Fletcher \cite{Fl76} and the more recent QMR (quasi minimum
residual) method of Freund and Nachtigal \cite{FN91}.  We show that
these Lanczos based methods can be very useful in carrying out the
extrapolation to the chiral limit: one can perform inversions on a
{\em whole} trajectory of masses simultaneously.  In other words, one
has to compute the propagator corresponding to the lightest mass only,
while all the heavier masses are almost given for free, at the price
of extra storage.

Moreover, we shall also point out how the particular symmetry
properties of the Wilson fermion matrix can be used to further reduce
the costs of each iterative step in QMR or BCG by a factor of 2. This
fact, mentioned in Ref.~\cite{FROMMER94} has recently attracted
attention in the lattice QCD community.

After discussing the basic properties of the Wilson fermion matrix in
Section~2, we will present the QMR and BCG algorithms in quite some
detail (Section~3) and explain the savings in computational effort due
to the structure of the Wilson fermion matrix.  Section 4 contains the
results of our numerical experiments on the CM5 parallel computer at
Wuppertal university. In particular, we will compare the QMR algorithm
with BiCGStab \cite{vV92,FHNLS94}, BCG and with the standard
over-relaxed MR (minimal residuum) method \cite{EISEN} on realistic
configurations for different values of $ \kappa $, approaching the
critical regime of very small relative quark masses. All these
calculations were done using standard odd-even preconditioning.

\section{Basic properties of the Wilson fermion matrix
\label{basic_properties_subsection}}

The Dirac operator in its discretized form as given by Wilson
\cite{WILSON} reads: 
\begin{equation} 
  M=\mbox{\bf 1} -\kappa D,
  \label{WILSON_1} 
\end{equation}
with the off-diagonal hopping term
\begin{equation} 
  D_{x,y} = \sum_{\mu=1}^4 (1-\gamma_{\mu})\,U_{\mu}(x)\, \delta_{x,y-\mu}+
  (1+\gamma_{\mu})\,U^{\dagger}_{\mu}(x-\mu)\, \delta_{x,y+\mu}.
  \label{WILSON_2} 
\end{equation}
In \eqnref{WILSON_2}, the $\{ U_{\mu}(x)\} $ represent the gauge
background field on a four-dimen\-si\-onal Euclidean space-time lattice.

In the following it is preferable to scale $M$ by a factor $\frac{1}{\kappa}$: 
$M \to \frac{1}{\kappa} M$. We shall thus 
consider the solution of the linear equation
\begin{equation} 
  M x = ( \frac{1}{\kappa} {\one} - D) x = \phi.
  \label{Wilson_eq}
\end{equation} 

The matrix $ M = \frac{1}{\kappa} \one - D $ in \eqnref{Wilson_eq} has
two important properties which will be crucial to algorithms for
solving \eqnref{Wilson_eq}:

\begin{itemize}
\item $ M $ is $ \gamma_5 $-symmetric, i.e.
$$ 
M \gamma_5 = \gamma_5 M^\dagger
$$
where $ \gamma_5 $ is the permutation matrix which commutes the Dirac components $ 1 $ with $ 3 $ and $ 2 $ with $ 4 $ on each lattice site. In particular, $ \gamma_5 $ is unitary and hermitian:
$$
\gamma_5 = \gamma_5^\dagger = \gamma_5^{-1} \ . 
$$
Multiplying a vector by $ \gamma_5 $ is a very cheap operation.

\item $ M $ is a {\em shifted matrix} with respect to its dependence on $ \kappa $, i.e. $ M $ is the sum of a multiple of the identity and a constant off-diagonal part $ D $.
\end{itemize}

When building up $ M $ one has the freedom to choose any ordering
scheme for the lattice sites. Subdividing the lattice in a
checkerboard style into even and odd sites and numbering all even
sites before the odd ones results in the following two-cyclic
structure for $ M $
\begin{equation}\label{Wilson_odd_even_eq}
M = \left( 
    \begin{array}{cc}
     (1/\kappa) \one & - D_{eo} \\
     -D_{oe}              & (1/\kappa) \one
     \end{array}
    \right)  .
\end{equation}
Using the indices `e' and `o' to denote even and odd sites, respectively, we can thus rewrite \eqnref{Wilson_eq} as
$$
\left(
    \begin{array}{cc}
     (1/\kappa) \one & - D_{eo} \\
     -D_{oe}              & (1/\kappa) \one
     \end{array}   
    \right)    
\left( \begin{array}{c} x_e \\ x_o \end{array} \right)
=
\left( \begin{array}{c} \phi_e \\ \phi_o \end{array} \right) 
$$
This equation separates into
\begin{equation}\label{prec_eq}
M_e x_e = \widetilde{\phi}_e \ ,
\end{equation}
\begin{equation}\label{xo_eq}
x_o = \kappa \cdot (\phi_o + D_{oe} x_e),
\end{equation}
where
\begin{equation}\label{Me_def_eq}
M_e = \frac{1}{\kappa^2} \one - D_{eo} D_{oe} \ ,
\end{equation}
\begin{equation}\label{new_phi_def_eq}
\widetilde{\phi}_e = \frac{1}{\kappa} \phi_e + D_{oe} \phi_0 \ . 
\end{equation}
\eqnref{prec_eq} is called the odd-even preconditioned system 
\cite{DEGRAND,GUPTA}, with $ M_e $
given by \eqnref{Me_def_eq}. It has become standard to solve the odd-even
preconditioned system rather than the original one since iterative methods for 
\eqnref{prec_eq} converge faster than for \eqnref{Wilson_eq}.

In our context, it is very important to notice that the matrix $ M_e $
of the preconditioned system conserves both basic properties of $ M $
as stated before, i.e.~$ M_e $ is still
$ \gamma_5 $-symmetric and a shifted matrix (with factor $ 1/\kappa^2
$ instead of $ 1/\kappa $). Therefore, using QMR or BCG we can take 
advantage of the particular structure of $M_e$ in just the same manner
as with  $M$. However, one should be aware of that the
new source term $ \widetilde{\phi}_e $ in \eqnref{new_phi_def_eq}
will depend on $ \kappa $ as soon as $ \phi_e \not= 0 $. This must be
accounted for when exploiting the shifted structure of $ M_e
$. Details will be given in the next section.  

\section{QMR and BCG}

We consider two different iterative methods for solving
\eqnref{Wilson_eq}: The BCG (biconjugate gradient) method of Fletcher
\cite{Fl76} and the QMR (quasi minimum residual) method of Freund and
Nachtigal \cite{FN91}. Generic formulations for these methods are
given in Algorithms \ref{BCG} and \ref{QMR}, see also
Ref.~\cite{Ba94}.  Although QMR looks somewhat more involved than BCG,
both methods require approximately the same computational work per
iteration: One matrix multiplication with $ M $ and another with $
M^\dagger $, the additional scalar and vector operations being
negligible in either method.  QMR will usually reduce the norm of the
residual in a much smoother manner than BCG does, thus making QMR more
stable numerically.

\begin{algor}[htb]
\begin{center}
\begin{minipage}{\textwidth}
\begin{tabbing}
\hspace{0cm} \= \hspace{2ex} \= \kill
\> choose $ x^0 \in \C^n $, set $ p^0 = r^0 = \phi - M x^0 $ \\
\> choose $ \widehat{r}^{0} \in \C^n $, set $ \widehat{p}^0 = \widehat{r}^0 $ \\\> for $ m = 0,1, \ldots $ \\
\> \> $ \delta_m = (\widehat{r}^m)^\dagger r^m / (\widehat{p}^m)^{\dagger} M p^m $ \\
\> \> $ x^{m+1} = x^m + \delta_m p^m $ \\
\> \> $ r^{m+1} = r^m - \delta_m M p^m $ \\
\> \> $ \widehat{r}^{m+1} = \widehat{r}^m - \overline{\delta}_m M^\dagger \widehat{p}^m $ \\
\> \> $ \rho_m = (\widehat{r}^{m+1})^\dagger r^{m+1} / (\widehat{r}^m)^\dagger r^m $ \\
\> \> $ p^{m+1} = r^{m+1} + \rho_m p^m $ \\
\> \> $ \widehat{p}^{m+1} = \widehat{r}^{m+1} + \overline{\rho}_m \widehat{p}^m $
\end{tabbing}
\end{minipage}
\end{center}
\caption[dummy]{BCG method.\label{BCG}}
\end{algor}
\begin{algor}
\begin{center}
\begin{minipage}{\textwidth}
\begin{tabbing}
\hspace{0cm} \= \hspace{2ex} \= \hspace{2ex} \= \kill
\> choose $ x^0 \in \C^n $, set $ \widetilde{v}^0 = b - M x^0 $ \\
\> choose $ \widetilde{w}^0 \in \C^n $ \\
\> \{initialize\} \\
\> set $ \mu_0 = \| \widetilde{v}^0 \|, \delta_0 = 1, c_{-1} = c_0 = 1, s_{-1} = s_0 = 0 \ ,$ \\
\> set $ p^{-1} = p^{-2} = v^{-1} = w^{-1} = 0 $ \\
\> for $ m = 0,1, \ldots $ \\
\> \> \{next Lanczos step\} \\
\> \> $ \rho_m = \| \widetilde{v}^m \|, \enspace \eta_m = \| \widetilde{w}^m \| $  \\
\> \> $ v^m = \widetilde{v}^m / \rho_m, w^m = \| \widetilde{w}^m \| \eta_m $
\\ \> \> $ \delta_m = (w^m)^\dagger v^m $ \\
\> \> $ \alpha_m = (w^m)^\dagger M v^m / \delta_m $ \\
\> \> $ \beta_m = \eta_m \delta_m / \delta_{m-1}, \gamma_m = \rho_m \delta_m /
      \delta_{m-1} $ \\
\> \> $ \widetilde{v}^{m+1} = M v^m - \alpha_m v^m - \beta_m v^{m-1} $ \\
\> \> $ \widetilde{w}^{m+1} = M^\dagger w^m - \overline{\alpha}_m w^m -
\overline{\gamma}_m w^{m-1} $ \\
\> \> \{update QMR recurrence coefficients\} \\
\> \> set $ \Theta_{m+1} = s_{m-1}  \beta_m, \widetilde{\varepsilon}_{m+1} = c_{m-1} \beta_m, \varepsilon_{m+1} = c_m \widetilde{\varepsilon}_{m+1} + s_m
\alpha_m, $ \\
\> \> \> $ \widetilde{\delta}_{m+1} = - \overline{s}_m \widetilde{\varepsilon}_{m+1} + c_m \alpha_m,  \nu_{m+1} =
(| \widetilde{\delta}_{m+1} |^2 + | \gamma_{m+1} |^2)^{1/2}, $ \\
\> \> \> $ c_{m+1} = | \widetilde{\delta}_{m+1} | / \nu_{m+1}, \overline{s}_{m+1} = 0 $ if $ \widetilde{\delta}_{m+1} = 0 $, \\
\> \> \> $ \overline{s}_{m+1} = c_{m+1} \gamma_{m+1} / \widetilde{\delta}_{m+1}$
if $ \widetilde{\delta}_{m+1} \not= 0 $, \\
\> \> \> $ \delta_{m+1} = c_{m+1} \widetilde{\delta}_{m+1} + s_{m+1} \gamma_{m+1} $ \\
\> \> \{update QMR iterate\} \\
\> \> set $ p^m = (v^m - \varepsilon_{m+1} p^{m-1} - \Theta_{m+1} p^{m-2}) / 
\delta_{m+1} $ \\
\> \> $ \widetilde{\mu}_m  = c_{m+1} \mu_m  $ \\
\> \> $ x^{m+1} = x^m + \widetilde{\mu}_m p^m $ \\
\> \> $ \mu_{m+1} = \overline{s}_{m+1} \mu_m $
\end{tabbing}
\end{minipage}
\end{center}  
\caption[dummy]{QMR method.\label{QMR}}
\end{algor}
The vectors $ \widehat{r}^0 $ in BCG and $ \widetilde{w}^0 $ in QMR can be chosen
freely. If we take $ \widehat{r}^0 = \widetilde{w}^0 $, it can be shown
\cite{Fr91} that $
r_{BCG}^m $ is a scalar multiple of $ \widetilde{v}^m $ and $ \widehat{r}^m $
is a scalar multiple of $ \widetilde{w}^m $. (We use the subscripts BCG and
QMR to distinguish between quantities which are otherwise denoted by the same
 symbol in
both methods). Thus, QMR and BCG are intimately related and one can show that
\cite{Fr91}
\begin{equation}\label{xBCG_eq}
x_{BCG}^m = x_{QMR}^m - \mu_m \frac{s_m}{c_m} p_{QMR}^{m-1}
\end{equation}
and
\begin{equation}\label{rBCG_norm_eq}
\| r_{BCG}^m \| = \| r^0 \|_2 \cdot |s_1 \ldots s_m | \cdot \frac{1}{c_m} \
.
\end{equation}

So, the BCG iterates can be retrieved very easily from the QMR iterates. The
generation of $ v^m $ and $ w^m $ as given in the QMR algorithm is called the
{\em non-symmetric Lanczos process}. This process is the common basis of QMR
and BCG.

Both methods, QMR and BCG can break down prematurely due to $\delta_m = 0$ in QMR
or zero divisors in $\rho_m$ or $\delta_m$ in BCG. This is why
the state-of-the-art package QMRPACK  \cite{FN94-2}
(distributed freely through the netlib server {\tt netlib@ornl.gov}) 
contains
modifications of QMR in which `look-ahead' Lanczos steps \cite{FGN93}
are incorporated. This
avoids premature breakdowns at the expense of extra storage and it further
enhances the numerical stability.

\begin{algor}[htb]
\begin{center}
\begin{minipage}{\textwidth}
\begin{tabbing}
\hspace{0cm} \= \hspace{2ex} \= \kill
\> choose $ x^0 \in \C^n $, set $ p^0 = r^0 = \phi - M x^0 $ \\
\> for $ m = 0,1, \ldots $ \\
\> \> $ \delta_m = (\gamma_5 r^m)^\dagger r^m / (A p^m)^\dagger (\gamma_5 p^m) $ \\
\> \> $ x^{m+1} = x^m + \delta_m p^m $ \\
\> \> $ r^{m+1} = r^m - \delta_m A p^m $ \\
\> \> $ \rho_m = (\gamma_5 r^{m+1})^\dagger r^{m+1} / (\gamma_5 r^m)^\dagger r^m $ \\
\> \> $ p^{m+1} = r^{m+1} + \rho_m p^m $
\end{tabbing}
\end{minipage}
\end{center}
\caption[dummy]{BCG exploiting the $ \gamma_5 $-symmetry.\label{BCG_GAMMA5}}
\end{algor}
If we choose $ \widehat{r}^0 = \gamma_5 r^0 $ in BCG, an easy calculation shows
that due to the $ \gamma_5$-symmetry of $ M $ we have $ \widehat{r}^m = \gamma_5
r^m $ for all $ m $ and the scalars $ \delta_m, \rho_m $ in BCG are all real.
Consequently, the computational effort per iteration reduces to only one matrix
multiplication, see Algorithm~\ref{BCG_GAMMA5}. Similarly, if we take $ \widetilde{w}^0 = \gamma_5
\widetilde{v}^0 $ in QMR, the non-symmetric Lanczos step yields $
\widetilde{w}^m = \gamma_5 \widetilde{v}^m $ for all $ m $ and the scalar
quantities in QMR again become all real. This simplified version of the Lanzcos
 process is given in Algorithm~\ref{QMR_GAMMA5}. So, once more, the multiplication with
$ M^\dagger $ can be saved, reducing the computational effort in QMR by a
factor of two. In the more general case of so-called 
$P$-symmetric matrices, the above simplifications
were already described in Ref.~\cite{FREUND93}
\begin{algor}[htb]
\begin{center}
\begin{minipage}{\textwidth}
\begin{tabbing}
\hspace{0cm} \= \hspace{2ex} \= \kill
\> choose $ \widetilde{v}^0 $ \\
\> set $ v^{-1} = 0, \quad \delta_0 = 1 $ \\
\> for $ m = 0,1, \ldots $ \\
\> \> $ \rho_m = \| \widetilde{v}^m \|$  \\
\> \> $ v^m = \widetilde{v}^m / \rho_m $ \\
\> \> $ \delta_m = (\gamma_5 v^m)^\dagger v^m $ \\
\> \> $ \alpha_m = (\gamma_5 v^m)^\dagger M v^m / \delta_m $ \\
\> \> $ \beta_m = \rho_m \delta_m / \delta_{m-1} $ \\
\> \> $ \widetilde{v}^{m+1} = M v^m - \alpha_m v^m - \beta_m v^{m-1} $
\end{tabbing}
\end{minipage}
\end{center} 
\caption[dummy]{The non-symmetric Lanczos process 
exploiting $\gamma_{5}$-symmetry.\label{QMR_GAMMA5}}
\end{algor}

Additionally, we note that $ M $ being a shifted matrix, $ M = \sigma \one
- D $ with $\sigma = 1/\kappa$ , we can rewrite the generation of $\tilde{v}_{m}$
in the non-symmetric Lanczos process of Algorithm~\ref{QMR_GAMMA5} as
$$
\widetilde{v}^m  =  - D v^{m-1} - (\alpha_{m-1} - \sigma) v^{m-1} -
\beta_{m-1} v^{m-2},
$$
with
$$
\alpha_{m-1} - \sigma = (\gamma_5v^{m-1})^{\dagger}(-D)v^{m-1}/\delta_{m-1}.
$$
This shows that the Lanzcos vectors $ \widetilde{v}^m$ 
and $ v^m$ depend only on $ D $ but not on $ \sigma $, provided one always has
the same
initial vector $ \widetilde{v}^0 $. The 
recurrence coefficients $ \beta_m, \eta_m $ remain unchanged  whereas
the coefficient $ \alpha_m $ changes to $ \alpha_m + \sigma $
if we change the matrix from $ -D $ to $ \sigma \one - D $.

Consequently, if we
simultaneously solve several systems
$$
( \frac{1}{\kappa_i} \one - D) x_i = \phi, \enspace i = 1, \ldots, l
$$
using QMR, the
Lanczos part itself has to be performed only once, provided we take the
initial guess $ x_i^0 = 0 $ for $ i = 1, \ldots, l $. In fact, 
$x_i^0 = 0 $ for $ i = 1, \ldots, l $ leads to the same initial residual
$ r_i^0 = \phi $ for $ i = 1, \ldots, l $ so that $ \widetilde{v}^0 = \phi $
represents the initial vector of the Lanczos process for all $ i $.
These observations go back to Ref.~\cite{FREUND93-2}.
 
Note that the odd-even preconditioned system \eqnref{prec_eq}
has the same shifted structure
as the original Wilson-Fermion matrix with $ \sigma = 1/\kappa^2 $. The right
hand side $ \widetilde{\phi}_e = \frac{1}{\kappa} \phi_e + D_{oe} \phi_0 $,
however, will usually depend on $ \kappa $ so that it is impossible to easily
find initial guesses which lead to the same initial residual. In that case we propose to consider the two systems
\begin{equation}\label{yz_eq}
\begin{array}{l}
M_e y_e = \phi_e\\
M_e z_e = D_{oe} \phi_0 \ ,
\end{array}
\end{equation}
which now both leave an initial residual that does not depend on $ \kappa $
if the initial guess is taken to be zero. Hence, again, for each of the two
systems the non-symmetric Lanczos process needs to be performed only once for
different values of $ \kappa $.

So, in the QMR method we can treat several values of $ \kappa $ at the same time
without introducing any additional matrix multiplications. This also holds for
the BCG iterates if we compute them from the QMR algorithm via \eqnref{xBCG_eq}.
The price to pay is 4 vectors extra storage for each additional 
value of $ \kappa $. This
can be reduced to 3 vectors if we refrain from updating the residuals. Indeed,
it was shown in Ref.~\cite{FN91} that
$$
\| r^m \| \leq  \| r^0 \|_2 \cdot \sqrt{m+1} \ |s_1, \cdots s_m| \
 =: \tau_m \ .
$$
The scalar $ \tau_m $ can be updated easily, and since it represents an upper
bound for $ \| r^m \| $ it can be used in a stopping criterion. It has turned
out to be a good choice checking for $ \tau_m \leq 10 \cdot \varepsilon $ if
one wants to have $ \| r^m \| \leq \varepsilon $, but, of course, the latter
inequality should be re-checked by explicitly computing $ r^m $ and $ \| r^m \| $ once $ \tau_m \leq 10 \cdot \varepsilon $ is fulfilled.

Finally, we just mention that all the above simplifications remain valid if we
incorporate the look-ahead Lanczos process \cite{FREUND93-2}.

\section{Results}

Our numerical computations were done on the CM5 parallel computer at
Wuppertal university. We tested and compared five different methods
for solving the odd-even preconditioned system \eqnref{prec_eq}: QMR
and BCG as described before, the standard over-relaxed MR (minimal
residual) method, the BiCGStab method and the usual conjugate gradient
method applied to the normal equation $$ M_e^{\dagger}M_e x_e =
M_e^{\dagger}\tilde{\phi}_e. $$ This method is abbreviated CGNE.

In our comparative study, we set high value on trying to be close to
realistic lattice gauge applications: at $\beta=6.0$, we have
generated an ensemble of $10$ decorrelated quenched gauge
configurations on a lattice of size of $16^4$, see also
Ref.~\cite{FHNLS94}.  We worked with a series of
$\kappa$-values, $\kappa=0.152, 0.153, 0.154, 0.155, 0.1553$, that
corresponds to a quark mass range $0.1 > m_q > 0.03$. In
Refs.~\cite{GUESKEN,GUPTA94,APE}, propagator computations on these
$\kappa$-values resulted in a nearly linear pion mass trajectory as
function of $1/\kappa$.

As is well known that the convergence behaviour of iterative solvers
can depend on the source vector, we adapted the Wuppertal source
smearing method \cite{SMEARING} to build our source: starting from a
location $x$ on a given time-slice the smeared source is generated
according to the smearing procedure
\begin{eqnarray}
\lefteqn{\phi(\vec x,t)\rightarrow\phi'(\vec x,t)=
\frac{1}{1-6\alpha}}\nonumber\\
&\times&
\Big\{
\phi(\vec x,t) + \alpha\sum_{i=1}^3\Big[
U_{i}(\vec x,t)\,\phi(\vec x+\vec e_i,t) +
U_{i}^{\dagger}(\vec x-\vec e_i,t)\,\phi(\vec x-\vec e_i,t)         
\Big]
\Big\},\nonumber\\
\end{eqnarray}
with the unit vector $\vec e_i$ pointing in spatial direction $i$.  We
used $\alpha=4$ and performed $100$ smearing iterations.

\begin{figure}
\begin{center}
\input{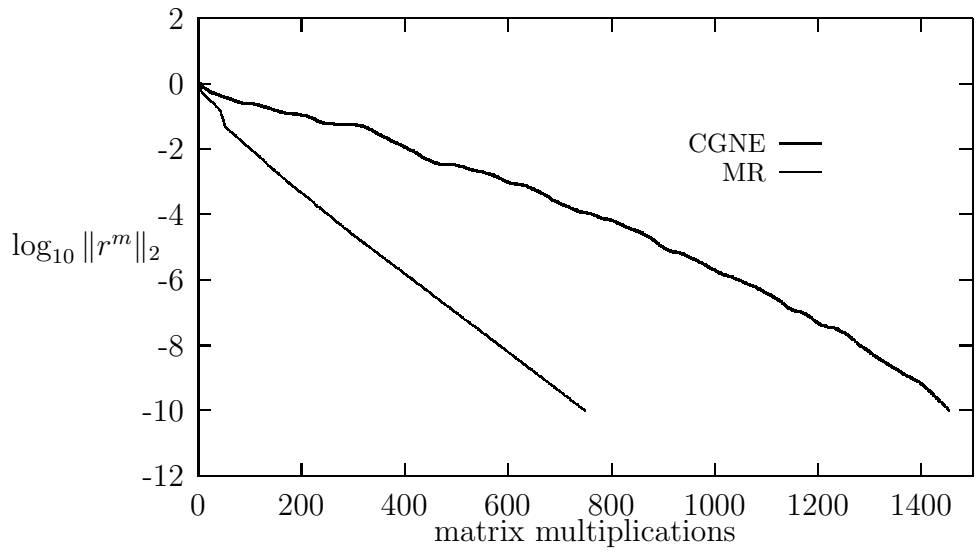}
\vspace*{1ex} 
\input{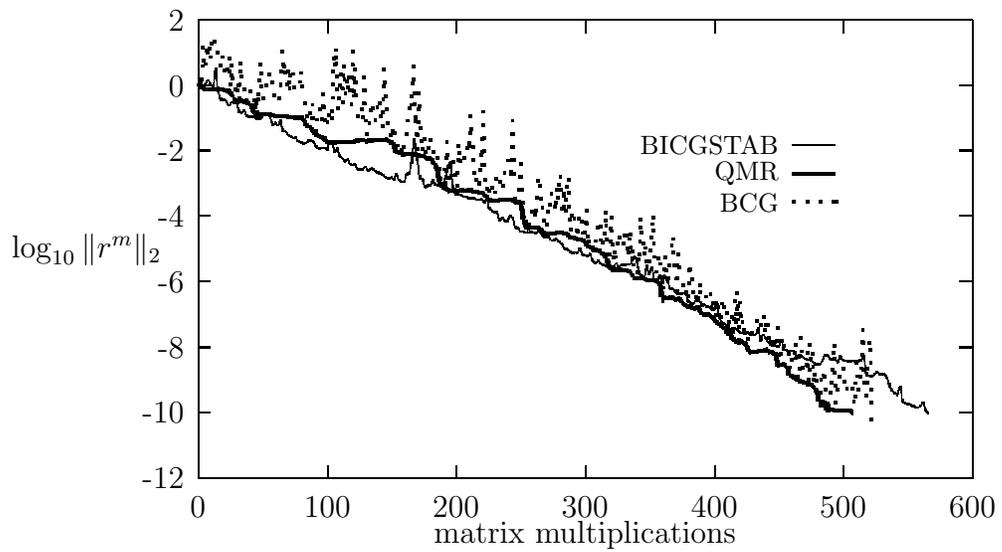}
\end{center}
\caption[dummy]{Convergence history for different methods\label{ONECONF}}
\end{figure}

Our first diagram (Figure~\ref{ONECONF}) reports result obtained on
one given configuration of the ensemble with a smeared source and
$\kappa = 0.155$. We display the convergence history for the different
methods by plotting the norm of the residual (normalized to $\|r^0\| =
1$) against the number of matrix multiplications involved.  Since in
either method all the computational effort is very highly concentrated
on the matrix multiplications, Figure~\ref{ONECONF} may also be
interpreted as giving the norm of the residuals as a function of
computing time.  Here, a matrix multiplication is a multiplication
with $M_e$ or $M_e^{\dagger}$. Note that each iterative step of CGNE
and BiCGStab requires two matrix multiplications, whereas MR, QMR and
BCG require only one since we take advantage of the
$\gamma_{5}$-symmetry.  In all methods, our starting vector was the
zero vector. In BiCGStab, the `shadow residual vector' $\hat{r}^0$ was
chosen equal to the inital residual, which means that we used BiCGStab
exactly as described in Ref.~\cite{FHNLS94}.

Figure~\ref{ONECONF} shows that CGNE is by far the slowest method and
that MR also performs substantially worse than the remaining three
methods, although we used the optimal relaxation factor in MR. This
factor was found to be 1.1 by numerical experimentation.
Figure~\ref{ONECONF} also illustrates the wide fluctuations in the
residual norm which one usually observes in BCG. On the other hand,
QMR converges very smoothly. The speed of convergence of all three
methods, QMR, BCG and BiCGStab looks quite comparable, with BiCGStab
being slightly better at the beginning, whereas BCG and QMR perform a
little better towards the end. To reach the required residual norm of
$10^{-10}$, QMR and BCG were the fastest of all methods with QMR
performing some 10\% better than BiCGStab. If we had not made use of
the $\gamma_{5}$-symmetry, QMR and BCG would have taken twice the
number of matrix multiplications so that both methods would become
clearly inferior to BiCGStab.  This is the reason why, based on
computations on a cold and a hot model configuration and without
exploiting $\gamma_5$-symmetry QMR and BCG were regarded as less
efficient than BiCGStab and its variant BiCGStab2 in Ref.~\cite{BF94}.

We now turn to demonstrate the additional savings possible by
exploiting the $\gamma_5$-symmetry together with the shifted structure
of the Wilson fermion matrix. We compare QMR-MULT, the QMR method for
solving \eqnref{prec_eq} om an entire set of $\kappa$-values
simultaneously (so the Lanczos process is carried out only once), with
a standard sequential computation on these $\kappa$ values. To make a
fair comparison, we used the educated guess technique in the latter
case, i.e.\ the final result of a computation for which the previous
$\kappa$ was taken as the new starting vector for the next $\kappa$.
For this serial treatment of the different values of $\kappa$ we tried
CGNE, over-relaxed MR, BiCGStab and QMR itself.  For these comparisons
we used the whole ensemble of $10$ configurations. The results for
CGNE were so inferior in comparison to the other methods that we
decided to not include them into our diagrams.

\begin{figure}[htb]
\centerline{\epsfxsize=10cm\epsfbox{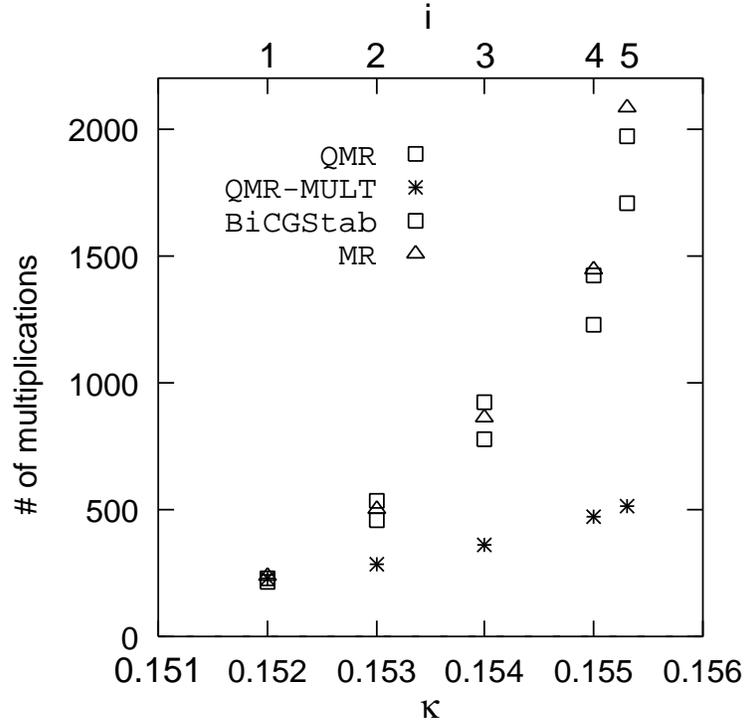}}
\caption[dummy]{Total number of matrix multiplications
as a function of the number $i$ of the $\kappa$'s for which the
calculations have been done.
\label{MULTS}}
\end{figure}

Figure~\ref{MULTS} gives the total number of matrix multiplications
$m$ as a function of the number of $\kappa$'s for which the
calculations have been done.
More precisely, a value of $i$ on the horizontal axis
refers to the calculation of $i$ different values $\kappa_1, \ldots ,
\kappa_i$, where
$$    
\kappa_1 = 0.152, \; \kappa_2 = 0.153, \;
\kappa_3 = 0.154, \; \kappa_4 = 0.155, \; \kappa_5 = 0.1553.
$$ In Figure~\ref{MULTS}, the source term was taken to be a point
source. The initial vector was the zero vector for all $\kappa$'s in
QMR-MULT and for the first value of $\kappa$ in all other methods. As
a consequence, a look-ahead Lanczos step had to be performed in QMR
and QMR-MULT at the very beginning of the iteration due to
$(\gamma_5\tilde{\phi}_e)^{\dagger}\tilde{\phi}_e = 0$.  We used a
simple modification of Algorithm~\ref{QMR} to perform this look-ahead
step\footnote{A more elaborate implementation of the QMR method
exploiting $\gamma_{5}$-symmetry and the shifted structure based on
QMRPACK is currently under development.}.

In either method the iteration was stopped when the norm of the
residual, weighted by the norm of the source term (which is the inital
residual if we take starting vector zero) was less than $10^{-10}$.
The representation in Figure~\ref{MULTS} shows the average values over
the whole sample of 10 configurations. The deviation of the results
for the individual configurations from the average was quite small,
ranging from less than 2\% for small values of $\kappa$ to never more
than 10\% for the largest $\kappa$ in either method.

\begin{figure}[htb]
\centerline{\epsfxsize=10cm\epsfbox{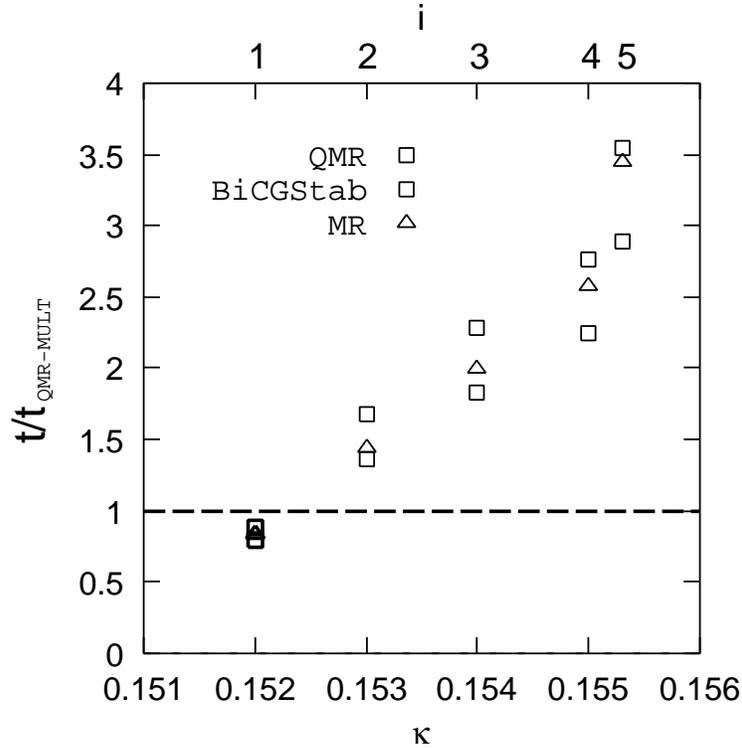}}
\caption[dummy]{
Computing time on the CM5 normalized to the time needed by QMR-MULT as
function of the number $i$ of $\kappa$'s for which the calculations
have been done.
\label{MULTS_KAP}}
\end{figure}

Figure~\ref{MULTS_KAP} refers to exactly the same computations as
Figure~\ref{MULTS}, showing now the computing time on the CM5 as
a function of the number of $\kappa$'s.
A comparison with Figure~\ref{MULTS} very clearly establishes that the
computing time is proportional to the number of matrix
multiplications.  Of course, in QMR-MULT, the additional work to be
spent in updating the iterates increases as we treat more $\kappa$'s
simultaneously, but this only minorly affects the overall computing
time.

Figure~\ref{MULTS} and \ref{MULTS_KAP} show that the more values of
$\kappa$ we treat simultaneously in QMR-MULT, the more we gain in
computing time against the educated guess variants. While for the case
of one single $\kappa$ all methods perform comparably well, the
`several-on-one-stroke' approach pays out as soon as we treat 2 or
more values of $\kappa$ simultaneously. For 5 values of $\kappa$,
QMR-MULT is almost three times as fast as the most rapid educated
guess method (via BiCGStab).

As was pointed out at the end of Section~2, using other sources than
point sources will usually require QMR-MULT to be performed on two
systems with different source terms, whereas this is not necessary for
the other methods. In that case we thus expect the QMR-MULT approach
to become approximately two times more slowly. Nevertheless, it would
be faster than all methods based on the educated guess as soon as we
treat 4 or more values of $\kappa$ simultaneously. For 5 values of
$\kappa$, for example, QMR-MULT will still be some 50\% better than
BiCGStab.

\section{SUMMARY}

In this note we have presented and tested an extension of the QMR
algorithm, the QMR-MULT, which exploits structural and symmetry
properties of the Wilson Fermion matrix in order to speed up the
inversions on a whole mass trajectory.  In the setting of a realistic
application, we have shown that the QMR-MULT can save a factor two to
four in computer time, compared to the fastest algorithms which are
presently in use.

\end{document}